\documentclass[doublecol,superscriptaddress,floats,floatfix,showpacs]{epl2}
\usepackage{multirow,amssymb,amsbsy,amsmath}
\usepackage{graphicx}
\usepackage{verbatim}
\usepackage{booktabs}
\usepackage{dcolumn}
\usepackage{bbm}
\makeatletter
\def\bra#1{\mathinner{\langle{#1}|}}
\def\ket#1{\mathinner{|{#1}\rangle}}

\DeclareMathOperator{\tr}{tr}

\usepackage{color}

\title{Efficient superdense coding in the presence of non-Markovian noise}

\author{Bi-Heng Liu\inst{1,2}\and
Xiao-Min Hu\inst{1,2} \and
Yun-Feng Huang\inst{1,2} \and
Chuan-Feng Li\inst{1,2}$\footnote{email: cfli@ustc.edu.cn}$ \and
Guang-Can Guo\inst{1,2} \and
Antti Karlsson\inst{3} \and
Elsi-Mari Laine\inst{3} \and
Sabrina Maniscalco\inst{3} \and
Chiara Macchiavello\inst{4} \and
Jyrki Piilo\inst{3}$\footnote{email: jyrki.piilo@utu.fi}$
}

\institute{
  \inst{1} Key Laboratory of Quantum Information, University of Science and Technology of China, CAS, Hefei, 230026, China\\
  \inst{2} Synergetic Innovation Center of Quantum Information and Quantum Physics, University of Science and Technology of China, CAS, Hefei, 230026, China\\
  \inst{3} Turku Centre for Quantum Physics, Department of Physics and Astronomy, University of Turku, FI-20014 Turun yliopisto, Finland \\
  \inst{4} Dipartimento di Fisica and INFN-Sezione di Pavia, via Bassi 6, 27100 Pavia, Italy
  }

\date{\today}

\abstract{
 Many quantum information tasks rely on entanglement, which is used as a resource,
for example, to enable efficient and secure communication. Typically, noise, accompanied by loss of entanglement, reduces the efficiency of quantum
protocols. We develop and demonstrate experimentally a superdense coding scheme with noise, where
the decrease of entanglement in Alice's encoding state does not reduce the efficiency of the information transmission.
Having almost fully dephased classical two-photon polarization state at the time of encoding with concurrence $0.163\pm0.007$, we reach values of mutual information close to $1.52\pm 0.02$ ($1.89\pm 0.05$) with 3-state  (4-state) encoding.
This high efficiency relies both on non-Markovian features, that 
Bob exploits just before his Bell-state measurement, and on 
very high visibility ($99.6\%\pm0.1\%$) of the Hong-Ou-Mandel interference within the experimental set-up.
Our proof-of-principle results with measurements on mutual information pave the way for exploiting non-Markovianity to improve the efficiency and security of quantum information processing tasks.
}

\pacs{03.65.Yz}{Decoherence; open systems; quantum statistical
methods}
\pacs{42.50.-p}{Quantum optics}
\pacs{03.67.-a}{Quantum information}

\begin{document}
\maketitle
Quantum information protocols exploit entanglement or other 
quantum resources~\cite{QI,IS}. 
Often, photons are used in communication~\cite{Gisin2007}, e.g., in superdense
coding (SDC)~\cite{Bennett,Peng,Mattle04, kwiat}. SDC is a paradigmatic protocol that employs entangled states in 
order to reach communication abilities that have no classical counterpart
and it is therefore at the heart of the opportunities offered by 
quantum information.
The original proposal~\cite{Bennett} provided a way to communicate 
two bits of classical information by having at disposal a maximally entangled 
two-qubit state, and by sending only a single qubit through a noiseless 
communication channel. More specifically, the sender (Alice), who shares a 
maximally 
entangled state (Bell state) with the receiver (Bob), encodes two bits of 
classical information by performing a unitary operation on her qubit (either the 
identity 
$\mathbbm{1}$ or one of the three Pauli operations $\sigma_x,\sigma_y,\sigma_z $), and then sends the qubit to Bob,
who will retrieve the encoded information by performing a Bell measurement
on the two-qubit system. 
The protocol was then generalised for an arbitrary entangled state $\rho_{AB}$
shared between Alice and Bob~\cite{hiroshima} and to many
users~\cite{PRL04}. In particular, the dense coding capacity for a
shared state $\rho_{AB}$, achieved by optimising the mutual information
between Alice and Bob over all possible encoding strategies and assuming 
a noiseless channel, was proven to take the simple form~\cite{hiroshima}
\begin{eqnarray}
C(\rho_{AB})=\log d+S(\rho_B)-S(\rho_{AB})\;, 
\label{Cnoiseless}
\end{eqnarray}
where $d$ is the dimension of Alice's system, 
$\rho_B$ is Bob's reduced density operator, and $S(\rho)=-\tr (\rho \log \rho) $ 
the von Neumann entropy. The protocol was also analysed in the context of 
noisy transmission channels~\cite{Bennett99,NJP10,chiara} and simple generalisations of the 
above expression were derived for the case of covariant noise.

Experimental implementations of superdense coding have been demonstrated with
photons~\cite{Mattle04,Peng,kwiat} and atoms~\cite{schaetz} over noiseless
channels, while the performance over a depolarising channel was reported
in~\cite{chiuri}. Since complete Bell state 
analysis is not possible within linear optical systems,
 the dense coding capacity is practically limited by the value $\log_2 3\approx 1.585$.
Higher values than the linear optical limit were reported in~\cite{kwiat},
where hyperentangled photons were employed.

In this Article we develop and experimentally realize a superdense coding
protocol with polarization entangled photons in a noisy environment.
Surprisingly, we show that noise can be tailored in order to obtain a 
performance 
close to the ideal case of maximally entangled states and no transmission
noise. 
To the best of our knowledge, we achieve the highest values of mutual 
information reported so far in the context of linear optics. 
Actually, we show that the mutual information of the protocol remains at a 
high constant value even if the entanglement of the two-qubit state, before 
Alice encodes the information, is reduced by noise acting on Alice's photon. 
Our results demonstrate that close-to-ideal superdense coding can be achieved 
even with arbitrarily small amount of entanglement in the degree of freedom 
into which the message is encoded.
The success of the protocol is based on the use of nonlocal memory 
effects~\cite{NLNM}, which are
induced by initial correlations between  the local environments of Alice's 
and Bob's qubits. In general, we refer to this type of environment as 
non-Markovian~\cite{wolf,NMprl,rivas,mani,NMNP,RevRMP, RivasRev}.
Recent theoretical results on open quantum systems strongly indicate that, 
under certain circumstances, non-Markovian noise is more beneficial with
respect to its Markovian counterpart for quantum information processing, 
metrology, and quantum 
communication~\cite{metrology,matsu,telep,channelcap}. 
While a full resource theory for non-Markovianity has not yet been developed, 
our results  provide the first experimental evidence,  to the best of our 
knowledge, on the utility of memory effects for quantum technologies.

Our SDC scheme and its experimental realization is based on a linear optical set-up where a polarization 
entangled pair of photons in the state $\ket{\psi(0)} =\ket{\Phi^+}= 
\frac{1}{\sqrt{2}}(\ket{HH} + \ket{VV})$
is generated by parametric downconversion. The scheme then consists of four 
main steps: 1) Local noise on Alice's photon 2) Alice's encoding 
3) Local noise on Bob's photon 4) Bell-state measurement.  
The polarization of the photons, which is used to encode the information, is coupled by quartz plates to their frequency distribution realizing a dephasing noise. Thereby in our scheme, the polarization degree of freedom plays the role of the open system and the frequency degree of freedom of the same physical object the role of the environment. Note that we do not have a proper heat bath in our set-up. However, we can control precisely the coupling between polarization and frequency and thereby introduce the noise in intentional way.
The Hamiltonian describing the local coupling between the polarization and frequency of each photon $j=A,B$ (Alice, Bob)
is 
\cite{NLNM}
\begin{equation}
H_j = \int \text{d} \omega_j \omega_j \big( n_V^j \ket{V} \bra{V} + n_H^j \ket{H} \bra{H}\big)\otimes \ket{\omega_j} \bra {\omega_j}. 
\label{Eq:H}
\end{equation}

Here $\omega_j$ is the frequency of photon $j$ and $n_V^j$ ($n_H^j$) the index of refraction of its polarization component $V$ ($H$).
We assume that $n_H^A - n_V^A = n_H^B - n_V^B\equiv\Delta n$.
The initial two-photon frequency state, in general, can be written as
$\int \text{d} \omega_A \text{d} \omega_B g(\omega_A, \omega_B) \ket{\omega_A}\ket{\omega_B}$ where
$g(\omega_A, \omega_B)$ is the joint probability amplitude and the corresponding joint probability distribution is $P(\omega_A,\omega_B) = |g(\omega_A, \omega_B)|^2$. 
We assume that the distribution $P(\omega_A,\omega_B)$ has a Gaussian form where the marginals have equal mean values $\langle \omega_A \rangle = \langle \omega_B \rangle = \omega_0 / 2$ and variances $C_{AA}= \langle \omega_A^2 \rangle- \langle \omega_A \rangle^2=C_{BB}$.
The correlation coefficient between the two frequencies is  $K=(\langle \omega_A \omega _B \rangle - \langle \omega_A\rangle \langle \omega_B\rangle ) /C_{\rm{AA}}$.
We note that using photons, the correlations between the frequencies, which eventually act as local environments, can be adjusted by controlling the pump in down conversion. If one considers other physical systems, e.g. atoms or ions for SDC, then correlating the uncontrolled ambient noise in Alice's and Bob's distant laboratories can be very challenging. However, the initial correlations between the local environments can be either quantum or classical 
 since the decoherence functions for the open system depend on the initial joint probability distribution of the environment~\mbox{\cite{expNLNM}}. 
Therefore, local operations and classical communication between Alice and Bob in their distant laboratories are sufficient to create the classical initial environmental correlations to exploit nonlocal memory effects and engineered noise for SDC.
Note also that quantum interference between reservoirs may open alternative possibilities for creating the required correlations~\cite{Chan2014}.

After the local noise on Alice's side, the polarization 
state shared between Alice and Bob is given by
 \begin{align}\label{decstate}
 \rho_{AB}(t_A) &= \frac{1}{2}\left[\ket{HH}\bra{HH} + 
\kappa_A(t_A)\ket{HH}\bra{VV} \right.\notag \\ 
 &+\left. \kappa_A^*(t_A)\ket{VV}\bra{HH}+\ket{VV}\bra{VV}\right],
\end{align}
where the decoherence function $\kappa_A$ as a function of Alice's interaction time $t_A$ is
\begin{align}
\label{kappaA}
\kappa_A(t_A) = \int \text{d} \omega_A \text{d} \omega_B e^{i t_A \omega_A\Delta n} |g(\omega_A, \omega_B)|^2.
\end{align}
State \eqref{decstate} with decoherece function \eqref{kappaA} from initial Bell-state $|\Phi^+\rangle$ is obtained by using Hamiltonian \eqref{Eq:H} on Alice's side and tracing out the frequency from the total system state
(see also references~\cite{NLNM,expNLNM}).
If we assume that there is no noise on Bob's side, the capacity of the 
protocol would be given by Eq.~(\ref{Cnoiseless}) with $ \rho_{AB}(t_A)$ 
given by the expression above, namely 
\begin{eqnarray}
C(\rho_{AB}(t_A))=2-H\left(\frac{1+|\kappa_A(t_A)|}{2}\right)\;, 
\label{C-Anoise}
\end{eqnarray}
where $H(x)=-x \log_2 x - (1-x) \log_2 (1-x)$ is the 
binary entropy function.
Alice applies now one of the local unitary operations $\left\{\mathbbm{1},
\sigma_x,\sigma_y,\sigma_z\right\}$ to encode her message to the decohered 
state (\ref{decstate}). 
Without loss of generality, let us assume that Alice applies $\sigma_x$.
In step 3 of our protocol Bob applies local noise to his qubit for the 
duration $t_B$. After this step the two-qubit state is 
\begin{align}
\rho_{AB}(t_A,t_B) &= \frac{1}{2}\left[\ket{VH}\bra{VH} + 
h(t_A,t_B)\ket{VH}\bra{HV} \right.\notag \\ 
 		&+ \left. h^*(t_A,t_B)\ket{HV}\bra{VH}+\ket{HV}\bra{HV}\right],
\end{align}
where the decoherence function, with $t_B=t_A$, is now
\begin{align}
h(t_A,t_A)=e^{i \omega_0 \Delta n t_A} e^{-C_{AA}t_A^2(1 + K)}.
\end{align}
In the ideal case of perfect anticorrelations in the 
frequency of 
the photons ($K=-1$), and if there does not exist other experimental imperfections, the magnitude of the decoherence function $h$ is equal 
to 1. Therefore, Bob, as a matter of fact, recreates a maximally entangled pair of 
photons. Despite of the presence of noise, the 
ideal capacity value equal to two becomes achievable.
The experimental results below show that apart from some minor frequency independent error sources we do practically achieve value $K=-1$.
However, for the sake of generality, we present first theoretical results valid for any value of $K$.
%  (for more details, see the supplementary materials \cite{supmat}).

Since noise on Bob's qubit acts locally, the SDC capacity of the present 
scheme can be computed as if it was applied before Alice's encoding and
with $t_A=t_B$ it is 
then given by
\begin{align}\label{MI}
C (\rho_{AB}(t_A) )=2-H\left(\frac{1+|\kappa_A(t_A)|^{2(1+K)}}{2}\right)\;. 
%\frac{1}{\ln(4)} \left\{ \left[ 1-|\kappa_A(t)|^{2(1+K)}\right] \ln\left[2-2|\kappa_A(t)|^{2(1+K)}\right] \right. \notag \\ 
%\left. + \left[1+|\kappa_A(t)|^{2(1+K)}\right] \ln\left[2+2|\kappa_A(t)|^{2(1+K)}\right]\right\} .
\end{align}
We can see that for high values of the correlation coefficient $K$ the 
SDC capacity in Eq.~\eqref{MI} exceeds notably the channel capacity of 
Eq.~\eqref{C-Anoise}, which corresponds to the case where Bob does not  
introduce any noise to his qubit. 
The increase in the capacity by the presence of noise on Bob's
qubit is due to nonlocal memory effects~\mbox{\cite{NLNM}}. 
In this scheme, despite the fact of having local interactions only, 
i.e. each qubit interacts with its own environment, the global two-qubit 
dynamical map is not a tensor product of the local maps. This is because 
the initial state of the composite environment contains correlations, as we 
have here, and as a consequence we can have, e.g., dynamics which is locally 
Markovian but globally non-Markovian. Therefore, we can make a connection 
between the correlation coefficient $K$, non-Markovianity $\mathcal{N}$ and 
the SDC capacity $C$, and demonstrate theoretically that non-Markovianity 
improves the information transmission. In detail, 
the amount of non-Markovianity $\mathcal{N}$ (as defined in Ref.~\cite{NMprl})
is related to the correlation coefficient $K$ and the decoherence function 
$\kappa_A$ in the following way $\mathcal{N} = |\kappa_A(t)|^{-K^2+1} - 
|\kappa_A(t)|$ (see also Ref.~\cite{telep}). 
By solving $|K|$ from this expression,
the SDC capacity $C$ [c.f.~Eq.\eqref{MI}] can be written explicitly as a function of non-Markovianity $\mathcal{N}$
\begin{align}
C (\mathcal{N}, |\kappa_A(t)|)= \nonumber \hspace{4.3cm}\\
 2-H\left(\frac{1+|\kappa_A(t)|^{2\left(1-\sqrt{1-\frac{\ln(\mathcal{N}+|\kappa_A(t)|)}{\ln(|\kappa_A(t)|)}}\right)}}{2}\right).
\end{align}
%For more details about about the calculation, the associated mutual information and a plot, see the supplementary material \cite{supmat}. 

%%%%%%%%%%%%% FIGURE 1 %%%%%%%%%%%%%%%%%%%%%%%%%%%%%%%%%%
\begin{figure}[t]
\centering
\includegraphics[width=0.45\textwidth]{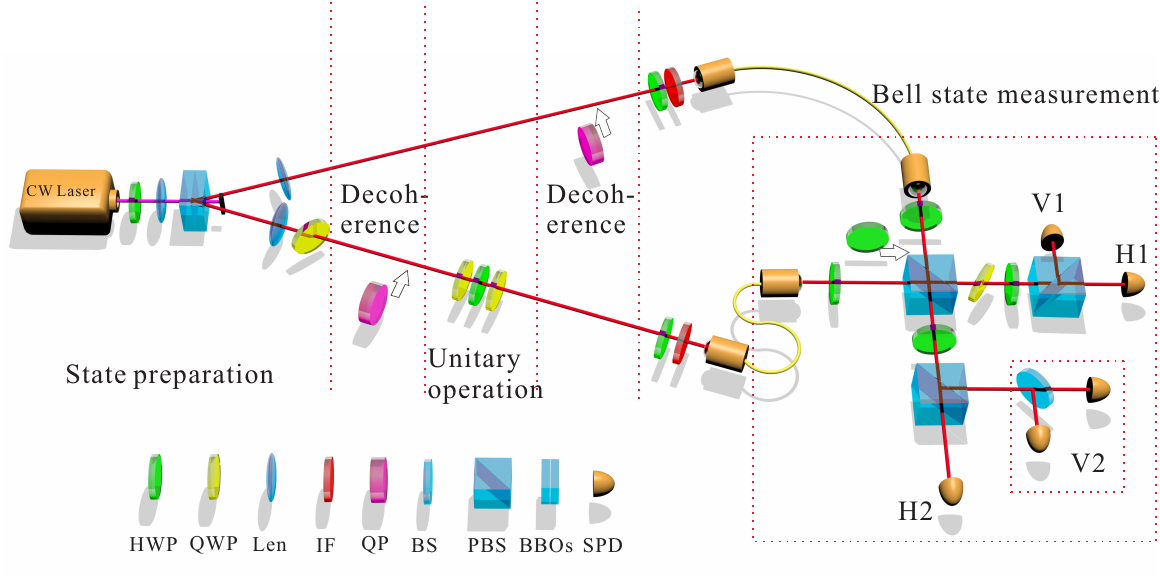}
\caption{\label{Fig:1} The experimental set-up. The abbreviations of the components are: HWP -- half wave plate, 
QWP -- quarter wave plate, Len -- lens, IF -- interference filter, QP -- quartz plate, BS-- beamsplitter, PBS --  polarizing beamsplitter, 
BBOs -- BBO crystals, and  SPD -- single photon detector.} 
\end{figure}
%%%%%%%%%%%%%%%%%%%%%%%%%%%%%%%%%%%%%%%%%%%%%%%%%%%%%%%%

%%%%%%%%%%%% FIGURE 2 %%%%%%%%%%%%%%%%%%%%%%%%%%%%%%%%%%
\begin{figure}[t]
\centering
\includegraphics[width=0.4\textwidth]{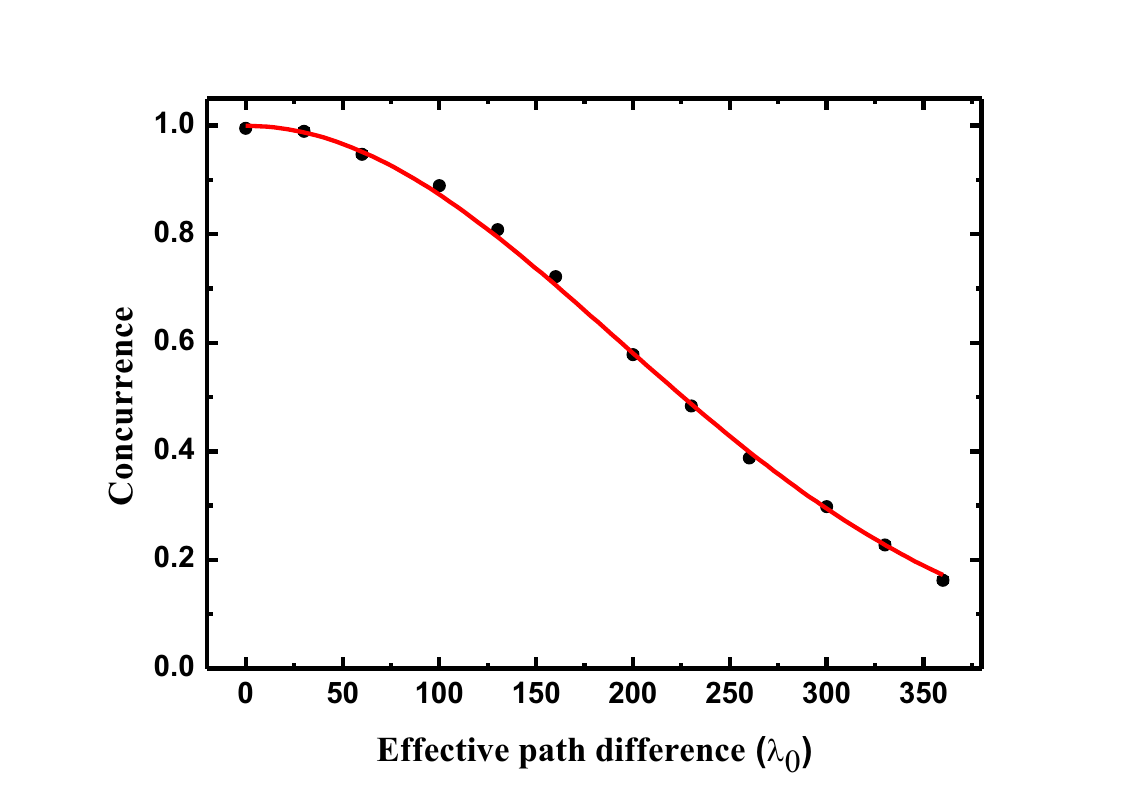}
\caption{\label{Fig:2} 
Entanglement vs the amount of Alice's noise. 
The points show the amount of entanglement, quantified by the concurrence, 
in the two-qubit state shared by Alice and Bob before Alice's encoding.
We use standard two qubit state tomography process~\mbox{\cite{James01}} to rebuild the density matrix of the two photon polarization state and then calculate the concurrence~\mbox{\cite{Wootters98}} from this density matrix.
The solid line is an exponential fit to the experimental result.
The error bars, which are due to the counting statistics and calculated using Monte Carlo simulation, are smaller than the marks of the data points.
}
\end{figure}
%%%%%%%%%%%%%%%%%%%%%%%%%%%%%%%%%%%%%%%%%%%%%%%%%%%%%%%%

%\subsection*{Experimental scheme}

\begin{figure}[t]
\centering
\includegraphics[width=0.4\textwidth]{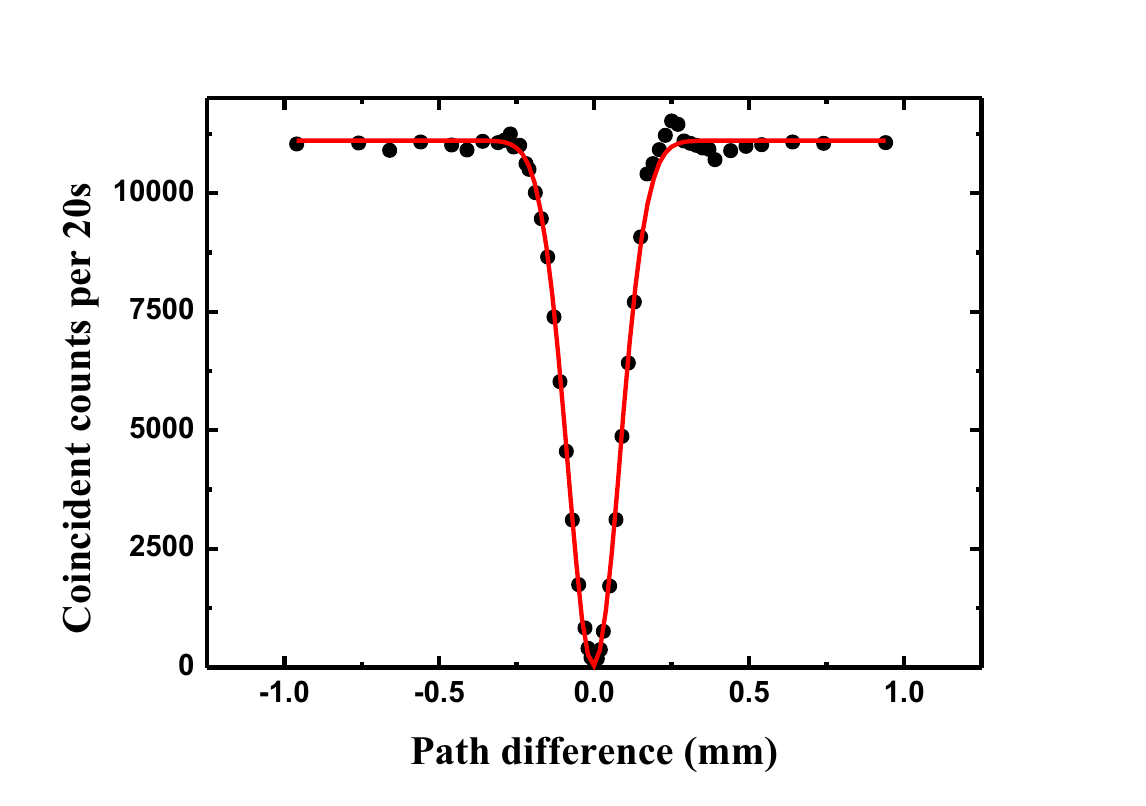}
\caption{\label{Fig:3} 
Hong-Ou-Mandel interference. We use single mode fiber to reduce the spatial mode mismatch. The visibility of the HOM interference in our experiment is $99.6\%\pm 0.1\%$. Red line is the Gaussian fit to the experimental data.}
\end{figure}

To realize the SDC protocol in the presence of noise, we use the experimental set-up displayed in Fig.~\ref{Fig:1}.
We use a continuous wave (CW) laser, with wavelength $\lambda_0=404~\textrm{nm}$. Compared to a pulsed laser pump down conversion source, the accidental coincidence rate is lower. In our case, the single photon count rate is about 10000 $\frac{1}{s}$, and the coincidence window  $3$ ns, which will cause an accidental coincidence rate about $0.3$ $\frac{1}{\textrm{s}}$. If one used pulsed laser (repetition rate about $76$ MHz), the accidental coincidence rate would be about $1.4$ $\frac{1}{\textrm{s}}$.The laser is focused onto two $0.3~\textrm{mm}$ thick type-I cut $\beta$-barium borate crystals to generate the two photon polarization entangled state 
$|\Phi ^+\rangle$~\cite{Kwiat99}.
One photon is sent to Alice and the other one is sent to Bob. On Alice's side, prior to her unitary encoding operation, local decoherence is implemented with quartz plates. 
We vary the amount of dephasing noise by controlling the quartz plate 
thickness.
Figure~\ref{Fig:2} shows the corresponding values of entanglement of the shared 
two-qubit state just before Alice performs her encoding operation.
Alice's polarization state encoding is realized with the sandwich of 
a quarter-wave plate, a half-wave plate and a quarter-wave plate. 
After this, Bob applies local noise to his photon by quartz plates and 
finally, after receiving Alice's photon, 
he performs the Bell-state measurement.
Our Bell-state measurement protocol uses three polarizing beamsplitters (see Fig.~\ref{Fig:1}) which are specially selected so that the extinction ratio is higher than 3000:1 on both sides. We also use single mode fibers to collect the photons and erase the spatial distinguishability of the photon pairs. To erase the spectral distinguishability of the two photons, we use two narrow-band interference filters for which the full width at half maximum is about 3nm.
With our polarizing beamsplitter set-up, we can distinguish $|\Phi^{\pm}\rangle=(|HH\rangle\pm|VV\rangle)/\sqrt{2}$. The state $|\Phi^{+}\rangle$  corresponds to the coincidence between H1 and H2,  or between V1 and V2. The state $|\Phi^{-}\rangle$, in turn, corresponds to the coincidence between H1 and V2,  or between V1 and H2. If we insert another HWP set at $45$ degree, then we convert $|\Psi^{\pm}\rangle=(|HV\rangle\pm|VH\rangle)/\sqrt{2}$ to $|\Phi^{\pm}\rangle$ and distinguish them.
This means that we can distinguish the four Bell-states in two 
measurement processes. However, if we have a photon number resolving 
detector we can directly distinguish the three Bell states in only one 
measurement process. Therefore, in the three state encoding experiment we can 
directly distinguish $|\Phi^+\rangle, |\Phi^-\rangle$, and $|\Psi^+\rangle$.
Experimentally we use a 50/50 beam splitter and two single photon detectors 
to replace V2~\cite{Mattle04}, 
see Fig.~\ref{Fig:1}. If two photons arrived at V2, 
then we can determine it with a possibility of $50\%$.
The key to the successful Bell-state measurement is the Hong-Ou-Mandel (HOM) 
interference. In our experiment the visibility of the HOM interference has a 
high value equal to $99.6\% \pm 0.1\%$, see Fig.~\ref{Fig:3}. 
The error bar is due to the counting statistics and calculated using Monte Carlo simulation.
Thus with our set-up, we observe nearly perfect HOM interference, which has previously  been observed only in a fiber beam splitter $(99.4\%\pm0.1\%)$ \cite{HOMvis}.

%%%%%%%%%%%% FIGURE 3 %%%%%%%%%%%%%%%%%%%%%%%%%%%%%%%%%%
%%%%%%%%%%%%%%%%%%%%%%%%%%%%%%%%%%%%%%%%%%%%%%%%%%%%%%%%

%\subsection*{Efficient information transmission}

From the Bell state measurement we can determine experimentally the mutual 
information between Alice and Bob with the noisy SDC scheme we use, by
using the explicit expression
\begin{align}\label{mutualinfo}
I(X:Y) = \sum_{x=1}^4 p_1(x)\sum_{y=1}^4 p(y|x) log_2 \frac{p(y|x)}{p_2(y)},
\end{align}
where $x$ and $y$ are the input and output variables corresponding in this 
case to the message encoded by Alice (one of the Bell states) and the 
measurement result obtained by Bob, with probability distributions $p_1(x)$ 
and $p_2(y)$ respectively, while 
$p(y|x)$ is the conditional probability of detecting the Bell state $y$ given
that the Bell state $x$ was transmitted.
We use our experimental scheme to implement both the 3-state protocol,
where information is encoded into the three states $|\Phi^+\rangle, 
|\Phi^-\rangle$ and $|\Psi^+\rangle$ with 
equal probabilities (i.e. $p_1(x)=1/3$), and the 4-state protocol that 
employs all the four Bell-states with $p_1(x)=1/4$.

%%%%%%%%%%%% FIGURE 4 %%%%%%%%%%%%%%%%%%%%%%%%%%%%%%%%%%
\begin{figure}[t]
\centering
\includegraphics[width=0.4\textwidth]{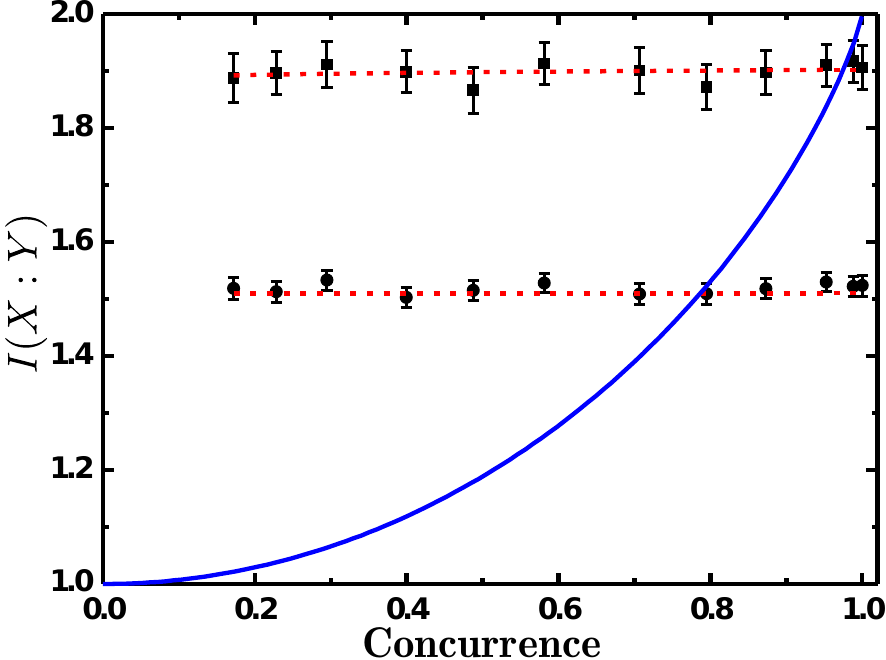}
\caption{
\label{Fig:4} 
Mutual information vs. the amount of concurrence.
We show the experimental results both for the 3-state encoding (black circles) and 4-state encoding (black squares).
The red dashed lines display the corresponding theoretical fits 
%(for more information, see the supplementary materials \cite{supmat})
and blue solid line the noiseless 4-state capacity from Eq.~(\ref{C-Anoise}) (no noise after encoding).
On the x-axis, concurrence is at the time of Alice's encoding and corresponds to y-axis of Fig.~2.
Note that this is also equal to the magnitude of the decoherence function $|\kappa_A|$~\mbox{\cite{NMNP}}.
Earlier experiments on SDC with photons have achieved the values of capacity 
1.13~\cite{Mattle04} and 1.63~\cite{kwiat}, both without any noise and latter by exploiting hyper entanglement while in Ref.~\cite{chiuri} the mutual information decreases when noise is applied.
The error bars are due to the counting statistics and calculated using Monte Carlo simulation.
}
\end{figure}
%%%%%%%%%%%%%%%%%%%%%%%%%%%%%%%%%%%%%%%%%%%%%%%%%%%%%%%%

Figure~\ref{Fig:4} shows the experimental results for the 3-state and 4-state encoding.
The plot displays the experimentally determined values of the mutual 
information compared with the theoretical predictions 
as function of the concurrence $C$ at the time of Alice's encoding.
Note that the amount of concurrence $C$ is also equal to the magnitude of the decoherence function 
$|\kappa_A|$ (see Sec.~II of Supplementary Information in~\mbox{\cite{NMNP}}).
For the theory, we give results in the absence  
and in the presence of noise on Bob's qubit.
With noise, the mutual information from Eq.\eqref{mutualinfo} can be written
for the 3-state encoding
\begin{align}
I(|\kappa_A(t)|,K,s)=  \hspace{8cm}\nonumber \\
 \frac{1}{\ln(8)}  
\Bigg[ 2 |\kappa_A(t)|^{2+2K}\text{arctanh}\left(|\kappa_A(t)|^{2+2K}\right)  \hspace{4cm}  \nonumber \\
 + \ln\left(-\frac{27}{4}\left(-1+|\kappa_A(t)|^{2+2K}\right)\right)
 +  \ln\left(1+|\kappa_A(t)|^{2+2K}\right) \Bigg] -s,   \hspace{1cm} 
\end{align}
and for the 4-state encoding
\begin{align}
I(|\kappa_A(t)|,K,s)=  \hspace{8cm}\notag \\
 \frac{1}{\ln(4)} 
 \Bigg[
\left(1-|\kappa_A(t)|^{2+2K}\right)\ln\left(2-2|\kappa_A(t)|^{2+2K}\right) \hspace{3.5cm}\notag \\
+\left(1+|\kappa_A(t)|^{2+2K}\right)\ln\left(2+2|\kappa_A(t)|^{2+2K}\right) \Bigg] - s.  \hspace{3cm}
\end{align}
Above, the fitting parameters are the correlation coefficient $K$ and $s$ which accounts for experimental imperfections.
These frequency independent imperfections are due to the experimental Bell-state preparation and Bell-state measurement.
In the least square fits, they obtain for 3-state encoding (4-state encoding) the values $K=-1.0$ and $s=0.0749$ ($K=-0.99995$ and $s=0.0975$). 

By looking at Fig.~\ref{Fig:4}, we observe that both in 3-state and in 4-state encoding, the measured mutual information is almost independent 
of the amount of noise introduced to the system. 
In other words, the reduction of entanglement in the polarisation degree of 
freedom at the time 
of encoding does not influence the efficiency of the information transmission. 
Therefore, it is possible to reach the values of mutual information  
$1.52\pm 0.02$ ($1.89\pm 0.05$) with 3-state (4-state) encoding
having concurrence in the state $\rho_{AB}(t_A)$, just before Alice's 
encoding, equal to 
$0.163 \pm 0.007$. Note that for the ideal case the
capacity with 3-state encoding is equal to $\log_2 3\approx 1.585$.
To the best of our knowledge, the above value of 
the experimental mutual information for the case of 3-state encoding is 
higher than all previously reported experimental values achieved in the context
of linear optical implementations or with trapped ions. The 
experimental points reported in Fig.~\ref{Fig:4} for the 4-state encoding represent a 
proof-of-principle demonstration of the efficiency of the 4-state protocol, since complete Bell
analysis is not available in our scheme, based on linear optical elements.
In our scheme, the high values of measured mutual information are based on nonlocal memory effects~\cite{NLNM, telep, expNLNM,entdis}.
The present results also show that it is indeed possible to implement local unitary operations 
between Alice's and Bob's local noise processes without disturbing the 
appearance and influence of the memory effects. 

In addition of the main results presented above -- almost ideal information transmission despite of noise -- 
the scheme also opens new possibilities to improve the security of the transmission. 
The decoherence may be used as a scytale cipher for quantum information. Let Alice and Bob agree beforehand on a common decoherence basis and the duration of the noise, which are unknown to Eve. Now, Alice adds noise to her qubit before sending it to Bob, thus making the message unreadable for Eve. Since Bob knows in which basis Alice's qubit decohered, he can utilise the nonlocal memory effects to recover the message by additional noise. Therefore, Bob's noise acts as the stick needed for decrypting the message and to read the scytale cipher. Note that Alice and Bob do not necessarily have to share the information about the correlation coefficient $K$ provided that it has high enough value as in the experiment demonstrated here.

It is also worth mentioning that the exploitation of nonlocal memory effects 
in the present scheme provides good efficiencies also when local
dephasing noise
is introduced on Alice's qubit after her encoding.
The efficiency of the 3-state coding scheme is not influenced by reordering of the noise
because dephased $|\psi^{\pm} \rangle \langle \psi^{\pm}  |$ states always have orthogonal support with respect to protected 
$|\phi^{\pm} \rangle \langle \phi^{\pm}  |$ states.
For 4-state coding the advantage with respect to
the corresponding Markovian scenario (i.e. $K=0$) is less striking 
than in the experimental situation presented above but it is nevertheless quite appreciable when the noise has finite duration.
For infinite duration of the noise, the 4-state case becomes equal to 3-state coding.
It is also worth mentioning that the nonlocal memory effects~\mbox{\cite{NLNM}}, which we exploit here, were originally discovered for dephasing noise. 
Even though dephasing is one of the most common decoherence mechanisms, it is an important open problem if and how the nonlocal memory effects can be generalized to other types of decoherence, e.g., depolarizing
 or dissipative noise. Therefore, if in the scheme 
above there are other noise sources in addition to dephasing, with current knowledge, this is expected to reduce the efficiency of the presented protocol.
Having a scheme which protects quantum properties simultaneously against all possible types of noise is such a grand task that it is out of the reach of the 
present results, both theoretical and experimental.
 
To conclude, we have demonstrated an efficient superdense coding scheme in the
presence of dephasing noise which provides almost ideal performance by exploiting 
nonlocal memory effects.
 As a matter of fact, we reach the almost ideal values of mutual information with arbitrary small amount of  entanglement in 
the degree of freedom used for the information encoding. 
To the best of our knowledge, we have also demonstrated experimentally, for the first time, 
that non-Markovian memory effects can be harnessed to improve the efficiency of quantum information protocols. 
\\

\acknowledgments
 The Hefei group acknowledges financial support from the National Basic Research Program of China (2011CB921200), the Strategic Priority Research Program (B) of the Chinese Academy of Sciences 
(Grant No.~XDB01030300), the National Natural Science Foundation of China (11274289, 11325419, 11374288, 11104261, 61327901, 61225025), and the Fundamental Research Funds for the Central Universities (WK2470000011). The Turku group acknowledges financial support from Magnus Ehrnrooth Foundation, and Jenny and Antti Wihuri Foundation. 
This work has been supported by EU Collaborative project QuProCS (Grant Agreement 641277).


\begin{thebibliography}{xx}


\bibitem{QI}
\Name{A. Galindo \and M. A. Mart\'{\i}n-Delgado}
%Information and computation: Classical and quantum aspects.
{\textit{Rev. Mod. Phys.}}, {\bf 74} (2002) 347.

\bibitem{IS}
Nature Insight - Quantum Coherence.
{\textit {Nature}}, {\bf 453} (2008) 1003.

\bibitem{Gisin2007}
\Name{N. Gisin \and R. Thew}
%Quantum communication.
{\textit{Nature Phot.}}, {\bf 1} (2007) 165.

\bibitem{Bennett} 
\Name{C. H. Bennett \and S. J. Wiesner}
%Communication via one- and two-particle operators on Einstein-Podolsky-Rosen states.
{\textit{Phys. Rev. Lett.}}, {\bf 69} (1992) 2881.

\bibitem{Peng} 
\Name{X. Li, Q. Pan, J. Jing, J. Zhang, C. Xie \and K. Peng}
%Quantum dense coding exploiting a bright Einstein-Podolsky-Rosen beam.
{\textit{Phys. Rev. Lett.}}, {\bf 88} (2002) 047904.

\bibitem{Mattle04}
\Name{ K. Mattle, H. Weinfurter, P. G. Kwiat \and A. Zeilinger}
 %Dense coding in experimental quantum communication.
 {\textit{Phys. Rev. Lett.}}, {\bf 76} (1996) 4656.

 \bibitem{kwiat}
\Name{J. T. Barreiro, T. C. Wei \and P. G. Kwiat}
%Beating the channel capacity limit for linear photonic superdense coding.
{\textit{Nature Phys.}}, {\bf 4} (2008) 282.

\bibitem{NLNM} 
\Name{E. M. Laine, H.-P. Breuer, J. Piilo, C.-F. Li \and C.C. Guo}
%Nonlocal memory effects in the dynamics of open quantum systems.
{\textit{Phys. Rev. Lett.}}, \textbf{108} (2012) 210402,
Erratum: {\textit{ibid.}} 111, 229901 (2013).

\bibitem{hiroshima} 
\Name{T. Hiroshima}
%Optimal dense coding with mixed state entanglement.
 {\textit{J. Phys. A: Math. Gen.}}, {\bf 34} (2001) 6907.

\bibitem{PRL04} 
\Name{D. Bruss, G. M. D'Ariano, M. Lewenstein, C. Macchiavello,
A. Sen(De) \and U. Sen} 
%Distributed quantum dense coding.
{\textit{Phys. Rev. Lett.}}, {\bf 93} (2004) 210501.

\bibitem{Bennett99}
\Name{C. H. Bennett, P. W. Shor, J.A. Smolin \and A. V. Thapliyal}
%Entanglement-assisted classical capacity of noisy quantum channels.
{\textit{Phys. Rev. Lett.}}, {\bf 83} (1999) 3081.

\bibitem{NJP10}
\Name{Z. Shadman, H. Kampermann, C. Macchiavello \and D. Bruss}
%Optimal super dense coding over noisy quantum channels. 
{\textit{New J. Phys.}},  {\bf 12} (2010) 073042.

\bibitem{chiara} 
\Name{Z. Shadman, H. Kampermann, C. Macchiavello \and D. Bruss}
%A review on super dense coding over covariant noisy channels.
 {\textit{Quantum Measurements and Quantum Metrology}}, {\bf 1} (2013)  21.

\bibitem{schaetz}

\Name{T. Schaetz, M. D. Barrett, D. Leibfried, J. Chiaverini, J. Britton, W. M. Itano, J. D. Jost, C. Langer \and D. J. Wineland}
%Quantum dense coding with atomic qubits.
 {\textit{Phys. Rev. Lett.}}, {\bf 93} (2004) 040505.

\bibitem{chiuri}
\Name{A. Chiuri, S. Giacomini, C. Macchiavello \and P. Mataloni}
%Experimental achievement of the entanglement-assisted capacity for the depolarizing channel.
 {\textit{Phys. Rev. A}}, {\bf 87} (2013) 022333.

\bibitem{wolf} 
\Name{M. M. Wolf, J. Eisert, T. S. Cubitt \and J. I. Cirac}
%Assessing non-Markovian quantum dynamics. 
{\textit{Phys. Rev. Lett.}}, {\bf 101} (2008) 150402.

\bibitem{NMprl} 
\Name{H.-P. Breuer, E.-M. Laine \and J. Piilo}
%Measure for the degree of non-Markovian behavior of quantum processes in open systems.
{\textit{Phys. Rev. Lett.}}, {\bf 103} (2009) 210401.

\bibitem{rivas} 
\Name{\'A. Rivas, S. F. Huelga \and M. B. Plenio}
%Entanglement and non-Markovianity of quantum evolutions. 
{\textit{Phys. Rev. Lett.}}, {\bf 105} (2010) 050403.

\bibitem{mani}
\Name{D. Chru\'sci\'nski \and S. Maniscalco}
%Degree of non-Markovianity of quantum evolution.
{\textit{Phys. Rev. Lett.}}, {\bf 112} (2014) 120404.

\bibitem{NMNP}
\Name{B.-H. Liu, L. Li, Y.-F. Huang, C. F. Li, G.-C. Guo, E.-M. Laine, H.-P. Breuer \and J. Piilo}
%Experimental control of the transition from Markovian to non-Markovian dynamics of
%open quantum systems.
{\textit{Nature Phys.}}, {\bf 7} (2011) 931.

\bibitem{RevRMP}
\Name{H.-P. Breuer, E.-M. Laine, J. Piilo \and B. Vacchini}
%Non-Markovian dynamics in open quantum systems,
{\textit{Rev. Mod. Phys.}}, {\bf 88} (2016) 021002.

\bibitem{RivasRev}
\Name{\'A. Rivas, S. F. Huelga \and M. B. Plenio}
%Quantum non-Markovianity: characterization, quantification and detection.
{\textit{Rep. Prog. Phys.}}, {\bf 77} (2014) 094001.

\bibitem{metrology}
\Name{A. W. Chin, S. F. Huelga \and M. B. Plenio}
%Quantum metrology in non-Markovian environments.
{\textit{Phys. Rev. Lett.}}, {\bf 109} (2012) 233601.

\bibitem{matsu}
\Name{Y. Matsuzaki, S. C. Benjamin \and J. Fitzsimons}
%Magnetic field sensing beyond the standard quantum limit under the effect of decoherence.
{\textit{Phys. Rev. A}}, {\bf 84} (2011) 012103.

\bibitem{telep} 
\Name{E.-M. Laine, H.-P. Breuer \and J. Piilo}
%Nonlocal memory effects allow perfect teleportation with mixed states.
{\textit{Sci. Rep.}}, \textbf{4} (2014) 4620.

\bibitem{channelcap}
\Name{B. Bylicka, D. Chru\'sci\'nski \and S. Maniscalco}
%Non-Markovianity and reservoir memory of quantum channels: a quantum information theory perspective.
{\textit{Sci. Rep.}}, \textbf{4} (2014) 5720.

\bibitem{expNLNM} 
\Name{B.-H. Liu, D.-Y. Cao, Y.-F. Huang, C.-F. Li, G.-C. Guo, E.-M. Laine, H.-P. Breuer \and J. Piilo}
%Photonic realization of nonlocal memory effects and non-Markovian quantum probes.
{\textit{Sci. Rep.}}, \textbf{3} (2013) 1781.

\bibitem{Chan2014}
\Name{C.-K. Chan, G.-D. Lin, S. F. Yelin \and M. D. Lukin}
 {\textit{Phys. Rev. A}}, {\textbf{89}} (2014) 042117. 

\bibitem{Kwiat99}
\Name{P. G. Kwiat, E. Waks, A. G. White, I. Appelbaum \and P. H. Eberhard}
%Ultrabright source of polarization-entangled photons.
 {\textit{Phys. Rev. A}}, {\textbf{60}} (1999) R773. 
 
 \bibitem{James01}
\Name{D. F. V. James, P. G. Kwiat, J. Munro \and A. G. White}
%Measurement of qubits.
 {\textit{Phys. Rev. A}}, {\textbf{64}} (2001) 052312.
 
 \bibitem{Wootters98}
\Name{W. K. Wootters}
% Entanglement of formation of an arbitrary state of two qubits.
 {\textit{Phys. Rev. Lett.}}, {\textbf{80}} (1998) 2245.
 
 \bibitem{HOMvis}
\Name{T. B. Pittman \and J. D. Franson}
%Violation of Bell's inequality with photons from independent sources.
{\textit{Phys. Rev. Lett.}}, {\bf 90} (2003) 240401.
 
\bibitem{entdis}
\Name{G.-Y. Xiang, Z.-B. Hou, C.-F. Li, G.-C. Guo, H.-P. Breuer, E.-M. Laine \and J. Piilo}
%Entanglement distribution in optical fibers assisted by nonlocal memory effects.
{\textit{EPL}}, \textbf{107} (2014) 54006.



\end{thebibliography}
\end{document}